\documentclass[%
 aip,
 amsmath,amssymb,
 reprint,%
]{revtex4-1}

\usepackage{graphicx}% Include figure files
\usepackage{dcolumn}% Align table columns on decimal point
\usepackage{bm}% bold math
\usepackage[utf8]{inputenc}
\usepackage[T1]{fontenc}
\usepackage{mathptmx}

\usepackage{upgreek}

\begin{document}

\preprint{AIP/123-QED}

\title[Propagation of spin waves through a Néel domain wall]{Propagation of spin waves through a Néel domain wall}
% Force line breaks with \\

\author{O. Wojewoda}
    \email{ondrej.wojewoda@vutbr.cz}
    \affiliation{Institute of Physical Engineering, Brno University of Technology, 616 69 Brno, Czech Republic}
\author{T. Hula}%
    \affiliation{Institute of Ion Beam Physics and Materials Research, D-01062 HZDR, Dresden, Germany}
\author{L. Flajšman}%
    \affiliation{CEITEC BUT, Brno University of Technology, 612 00 Brno, Czech Republic}
\author{M. Vaňatka}%
    \affiliation{CEITEC BUT, Brno University of Technology, 612 00 Brno, Czech Republic}
\author{J. Gloss}%
    \affiliation{Institute of Applied Physics, TU Wien, 1040 Vienna, Austria}
\author{J. Holobrádek}%
    \affiliation{Institute of Physical Engineering, Brno University of Technology, 616 69 Brno, Czech Republic}
\author{M. Staňo}%
    \affiliation{CEITEC BUT, Brno University of Technology, 612 00 Brno, Czech Republic}
\author{S. Stienen}%
    \affiliation{Institute of Ion Beam Physics and Materials Research, D-01062 HZDR, Dresden, Germany}
\author{L. Körber}%
    \affiliation{Institute of Ion Beam Physics and Materials Research, D-01062 HZDR, Dresden, Germany}
    \affiliation{Fakultät Physik, Technische Universit\"at Dresden, D-01062 Dresden, Germany}
\author{K. Schultheiß}%
    \affiliation{Institute of Ion Beam Physics and Materials Research, D-01062 HZDR, Dresden, Germany}
\author{M. Schmid}%
    \affiliation{Institute of Applied Physics, TU Wien, 1040 Vienna, Austria}
\author{H. Schultheiß}%
    \affiliation{Institute of Ion Beam Physics and Materials Research, D-01062 HZDR, Dresden, Germany}
    \affiliation{Fakultät Physik, Technische Universit\"at Dresden, D-01062 Dresden, Germany}
\author{M. Urbánek}
    \email{michal.urbanek@ceitec.vutbr.cz}
    \affiliation{Institute of Physical Engineering, Brno University of Technology, 616 69 Brno, Czech Republic}
    \affiliation{CEITEC BUT, Brno University of Technology, 612 00 Brno, Czech Republic}

\date{\today}% It is always \today, today,
             %  but any date may be explicitly specified

\begin{abstract}
Spin waves have the potential to be used  as a new platform for data transfer and processing as they can reach wavelengths in the nanometer range and frequencies in the terahertz range. To realize a spin-wave device, it is essential to be able to manipulate the amplitude as well as the phase of spin waves. Several theoretical and recently also experimental works have shown that the spin-wave phase can be manipulated by the transmission through a domain wall (DW). Here, we study propagation of spin waves through a DW by means of micro-focused Brillouin light scattering microscopy ($\mathrm{\upmu}$BLS). The acquired 2D spin-wave intensity maps reveal that spin-wave transmission through a Néel DW is influenced by a topologically enforced circular Bloch line in the DW center and that the propagation regime depends on the spin-wave frequency. In the first regime, two spin-wave beams propagating around the circular Bloch line are formed, whereas in the second regime, spin waves propagate in a single central beam through the circular Bloch line. Phase-resolved $\mathrm{\upmu}$BLS measurements reveal a phase shift upon transmission through the domain wall for both regimes. Micromagnetic modelling of the transmitted spin waves unveils a distortion of their phase fronts which needs to be taken into account when interpreting the measurements and designing potential devices. Moreover, we show, by means of micromagnetic simulations, that an external magnetic field can be used to move the circular Bloch line within the DW and to manipulate spin-wave propagation. 
\end{abstract}

\maketitle

\begin{figure*}
\includegraphics{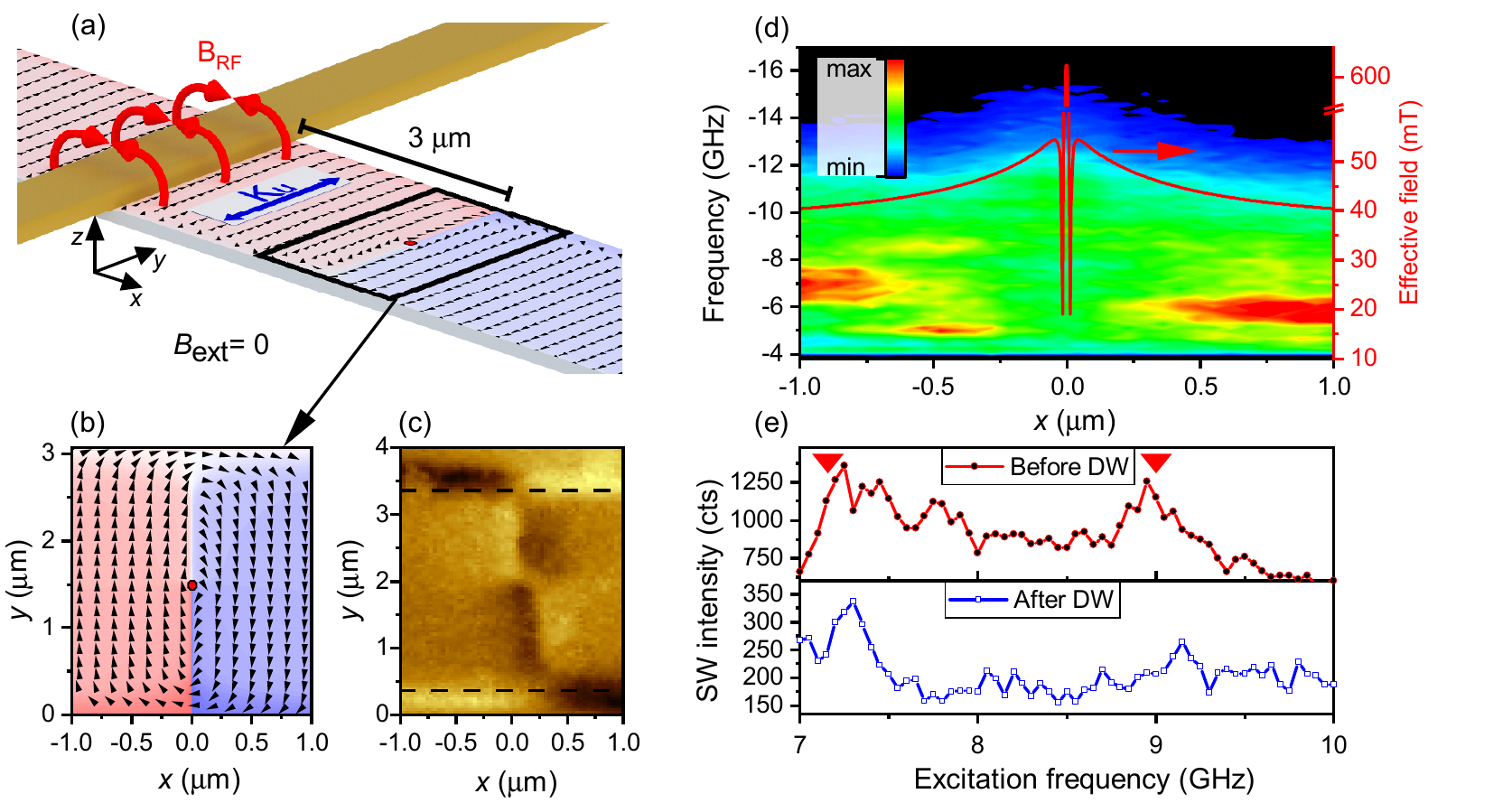}% Here is how to import EPS art
\caption{(a) Scheme of the experimental geometry. The ocher wire represents the excitation antenna producing the RF magnetic field. The uniaxial anisotropy (easy axis) created by the FIB irradiation is shown as a blue double arrow. The domain wall is positioned $3\,\upmu$m from the edge of the antenna. (b) Micromagnetic simulation of the DW state. (c) Magnetic force microscopy image of the DW. The dashed lines indicate the borders of the waveguide. (d) Plot of the BLS thermal spectra acquired while scanning the laser spot across the DW (located at $x=0\,\upmu\mathrm{m}$).The color code represents the spin-wave intensity and the red curve (right axis) represents the magnitude of the local effective field. (e) Frequency sweeps for positions $x=-0.5\,\upmu\mathrm{m}$ (before the DW) and $x=0.5\,\upmu$m (after the DW). The red arrows indicate prominent frequencies used in following experiments.} 
\end{figure*}
%Other aplications of DWs, DWs could be used as sw conduits, Albisetti DW
Magnonics offers a new promising concept for devices with the potential to go beyond CMOS technology in cases when low power computing is desired as Joule losses are effectively circumvented by the nature of the operation of magnonic devices. It was shown that it is possible to fabricate an all magnon transistor\cite{CHUMAK14} and spin-wave logic gates based on a Mach-Zehnder interferometer\cite{SCHNEIDER08, KOSTYLEV05, LEE08, VASILIEV07}. These gates rely on a controlled phase shift of spin waves, which might be achieved, e.g., by magnetic domain walls\cite{HERTEL04, BAYER05, AU12, HOIKI20}.

For this reason, the propagation of spin waves through domain walls (DW) is of high interest. Recently, it has been shown that domain walls can serve as a spin-wave valve depending on its magnetic state (head-to-head and tail-to-tail spin configurations)\cite{HAMALAINEN18}. In samples with perpendicular magnetic anisotropy, it was demonstrated, that DWs can affect both spin-wave amplitude and phase and that spin waves can move the DW by spin torque phenomena\cite{HAN19, HAN09, YAN11, SEO11}. Moreover, DWs may also serve as a source and probe of spin waves with short wavelengths\cite{WOO17} or can be used as magnonic conduits by themselves \cite{WAGNER16, ALBISETTI18}. However, there was no experimental study on the spatial evolution of spin waves transmitted through a DW in a magnonic waveguide. A thorough understanding of the behavior of spin waves in magnonic waveguides hosting DWs is necessary for designing future integrated magnonic circuits as they present the most elementary building block of magnonic circuitry.

In this study, we focused on spatial imaging of spin-wave propagation through a Néel domain wall confined in an in-plane magnetized magnonic waveguide. To prepare a sample suitable for this study, we used a metastable fcc $\mathrm{Fe}_{78}\mathrm{Ni}_{22}$ thin film with a nominal thickness of $16\,\mathrm{nm}$. The films were grown on a Cu(001) single crystal substrate under UHV conditions by the evaporation from $\mathrm{Fe}_{78}\mathrm{Ni}_{22}$ rod ($2\,\mathrm{mm}$ thick, purity 99.99\%, heated by electron bombardment) at a residual CO pressure of $5\times10^{-10}\,\mathrm{mbar}$ \cite{GLOSS13}. After deposition, the sample was transferred to the FIB-SEM microscope for FIB irradiation. We oriented the long axis of the waveguides along the fcc[010] direction of the Cu substrate in order to obtain the highest possible magnetocrystalline anisotropy \cite{URBANEK18}. To imprint the anisotropy direction perpendicularly to the long axis of the waveguide, we wrote the structures with a single pass of a 30\,keV Ga$^+$ ion beam (30\,nm spot, beam current of 150\,pA and 5\,$\upmu$s dwell time) with the fast scanning direction rotated by 80$^\circ$ from the waveguide’s long axis. The resulting ion dose of 4$\times$10$^{15}$\,cm$^{-2}$ has given reliable growth conditions for all the waveguides.

%The bcc Fe$_{78}$Ni$_{22}$ magnetic waveguides with dimensions $3\times20\,\upmu\mathrm{m}^2$ were written into the non-magnetic fcc layer by focused ion beam (Tescan Lyra3 FIB/SEM) induced phase transformation (fcc $\rightarrow$ bcc) using the procedure described in\cite{FLAJSMAN19}. 
%The bcc Fe$_{78}$Ni$_{22}$ magnetic waveguides with the dimensions of $3\times20\,\upmu\mathrm{m}^2$ were written into the non-magnetic fcc layer by inducing the fcc $\to$ bcc phase transformation with a focused ion beam (Tescan Lyra3 FIB/SEM), using the procedure described in\cite{FLAJSMAN19}. 
The produced waveguides had an effective anisotropy field of 40$\,$mT with an in-plane easy axis perpendicular to the waveguide's long axis [see Fig. 1(a)].
The effective anisotropy field is the result of the two competing contributions – the uniaxial magnetocrystalline anisotropy coming from the FIB writing process and the shape anisotropy having its origin in the waveguide's demagnetizing field.
This configuration allows for spin wave propagation in Damon-Eshbach (DE) geometry in zero external magnetic field and for stabilization of a domain wall in the vicinity of a spin-wave source. Spin waves were excited by a stripline antenna placed on top of the waveguide [see Fig. 1(a)]. The 1 $\upmu$m wide antenna, made of a multilayer structure consisting of a $20\,\mathrm{nm}$ $\mathrm{SiO}_2$ insulating layer, $5\,\mathrm{nm}$ Ti, $85\,\mathrm{nm}$ Cu and a $10\,\mathrm{nm}$ Au capping layer was patterned by electron beam lithography using PMMA resist, followed by e-beam evaporation of the layers and subsequent lift-off process. The desired two-domain state of the waveguide was stabilized by applying an external magnetic field in $+y$ direction while observing the magnetization by a Kerr microscope (evico magnetics GmbH). The DW was positioned at a distance of approximately 3$\,\upmu$m from the antenna in order to obtain a sufficient BLS signal on both sides of the domain wall. The position of the domain wall was stable in zero external magnetic field with a depinning field of $1.0\pm0.1\,$mT (obtained by Kerr microscopy experiments). 

To further reveal the internal structure of the DW, we performed magnetic force microscopy measurements (Bruker Dimension Icon, ASYMFMLM probe tip), supplemented by micromagnetic simulations using the MuMax3 software\cite{VANSTEENKISTE14} and the following parameters: simulation area $3\times16\,\mathrm{\upmu m^2}$, cell size $4\times4\times16\,\mathrm{nm^3}$, $M\mathrm{_s} = 1409\,\mathrm{kA/m}$, $K\mathrm{_u} = 28.9\,\mathrm{kJ/m^3}$, $A_\mathrm{ex}=11\,\mathrm{pJ/m}$. These material parameters were obtained from fitting the spin-wave dispersion measured by phase-resolved BLS using the model of Kalinikos-Slavin\cite{KALINIKOS86} as described in\cite{FLAJSMAN19}. We also confirmed that the simulation with twice smaller cell size gives identical result. From the results shown in Figs. 1(b) and 1(c) it is apparent that the domain wall is a symmetric Néel wall with a circular Bloch line (vortex) in the middle\cite{HUBERT98, albisetti2018stabilization}. The Bloch line is enforced by the surrounding topology (antiparallel magnetization along the two edges). The occurrence of the Bloch line results in a better dipolar energy minimization between the charges of the wall and those on the sides of the strip.  

Spin-waves propagation was studied using micro-focused Brillouin light scattering ($\mathrm{\upmu}$BLS)\cite{SEBASTIAN15}. In order to get better insight into internal magnetic field structure, we measured thermally excited spin waves while scanning the $\mathrm{\upmu}$BLS spot ($300\,\mathrm{nm}$ in diameter) in the center of the waveguide, across the DW [Fig. 1(d)]. The total length of the line scan was $2\,\mathrm{\upmu m}$, with the spatial step of 50$\,$nm and the acquisition time for a single spectrum was $240\,\mathrm{s}$. We observe an increase of spin-wave frequencies by approximately $2\,\mathrm{GHz}$ in the DW region (the DW is located at $x=0\,\upmu\mathrm{m}$) in comparison to the uniformly magnetized area. This increase can be attributed to approximately $15\,\mathrm{mT}$ increase in the local effective magnetic field. The effective magnetic field obtained from micromagnetic simulation is plotted together with the thermal spectra [Fig. 1(d), red curve]. Note, that at the exact position of the DW, the effective field exceeds $600\,\mathrm{mT}$ but this peak is spatially restricted only to a region of approximately $10\,\mathrm{nm}$ (an order of magnitude below the $\mathrm{\upmu}$BLS spatial resolution), and its contribution is not present in the acquired spectra.

\begin{figure*}
\includegraphics{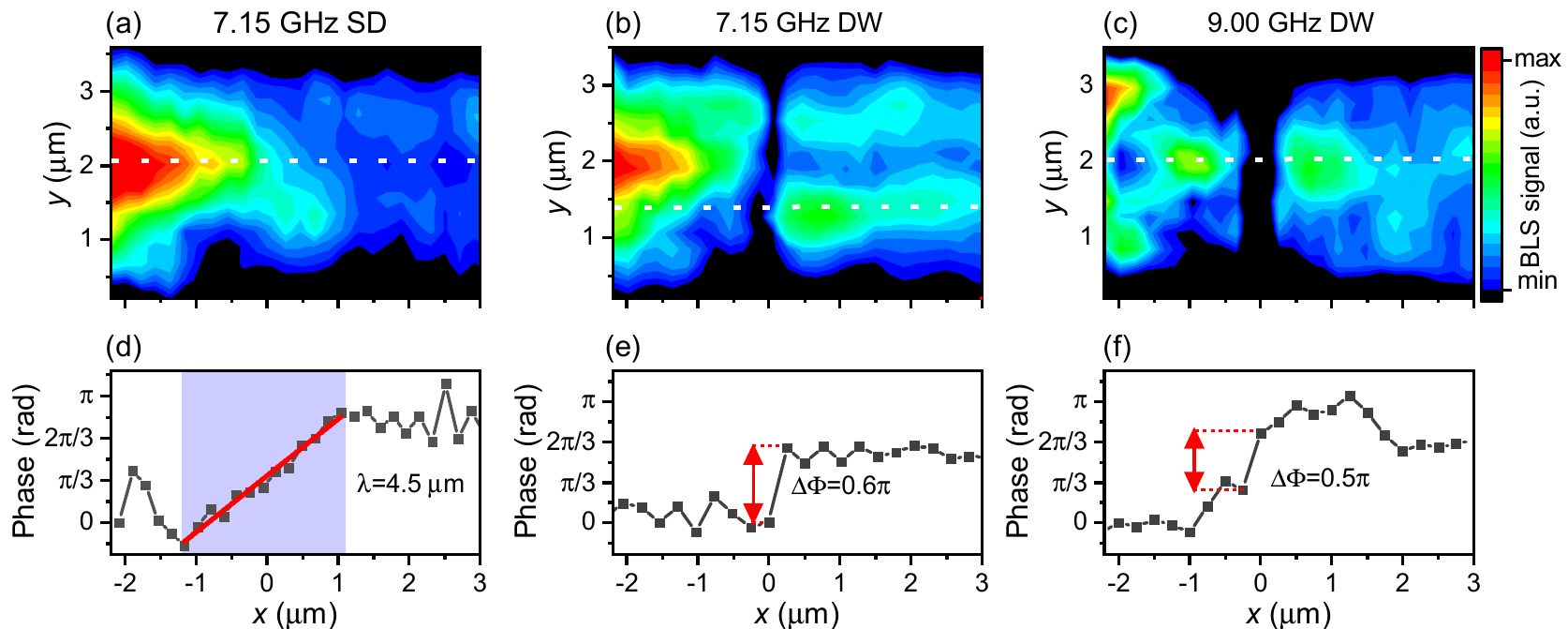}% Here is how to import EPS art
\caption{Spin-wave intensity maps for excitation frequency of $7.15\,\mathrm{GHz}$ (a) in single domain (SD) state and (b) in two-domain state separated by the Néel domain wall (DW state) and (c) for excitation frequency of $9.00\,\mathrm{GHz}$ in DW state. The white dashed lines indicate the positions of measured phase. (d) Evolution of the spin-wave phase along the white dashed line at the frequency of $7.15\,\mathrm{GHz}$ in SD state and (e) DW state. (f) Spin-wave phase for excitation frequency of $9.00\,\mathrm{GHz}$ in DW state. The blue region in (d) indicates a region of detectable linear phase evolution. The red line represents the linear fit of this data and the spin-wave wavelength is calculated from the slope of the fitted line.}
\end{figure*}

\begin{figure*}[t]
\includegraphics{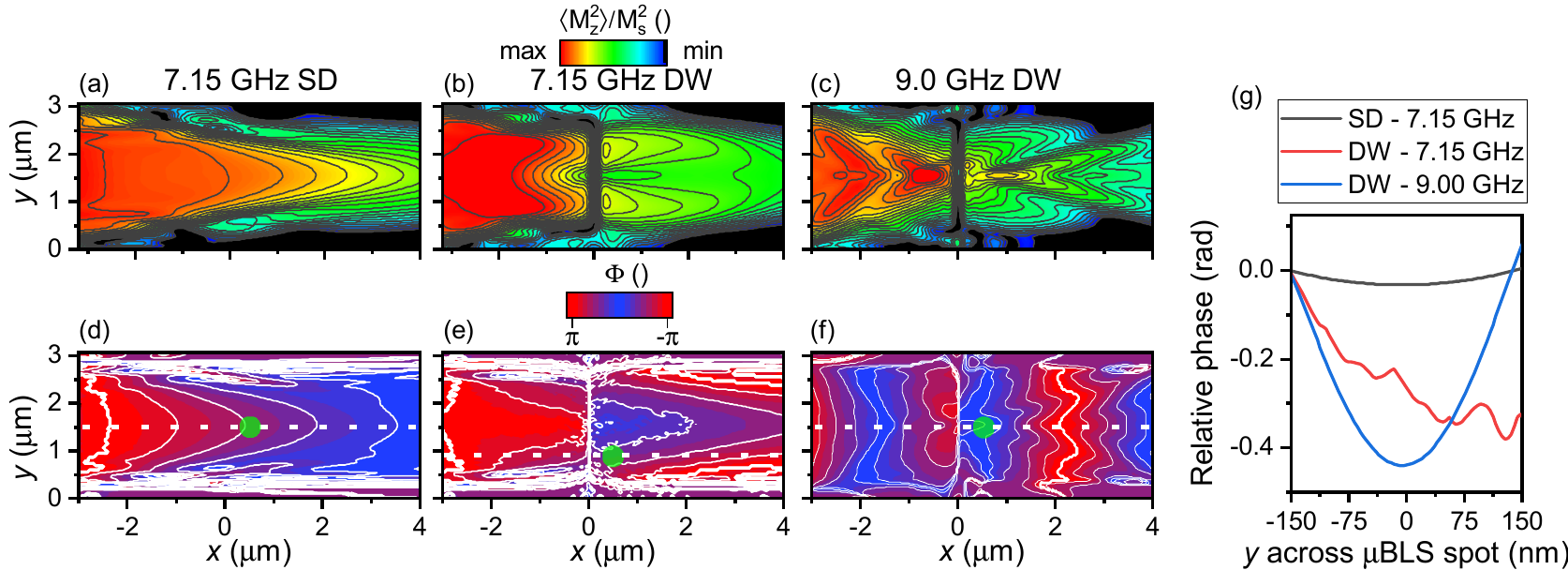}% Here is how to import EPS art
\caption{(a)-(c) Simulated 2D maps of the squared out-of-plane magnetization component averaged over two periods of excitation. (a) SD state, (b) DW state excited at $7.15\,\mathrm{GHz}$ and (c) DW state excited at $9.00\,\mathrm{GHz}$.(d)-(f) Phase maps corresponding to the respective intensity maps. (d) SD state, (e) DW state excited at $7.15\,\mathrm{GHz}$ and (f) DW state excited at $9.00\,\mathrm{GHz}$. The white dashed lines indicate the positions where the phase was experimentally measured. The green dots represent $\mathrm{\upmu}$BLS spots at which the simulated relative phase is evaluated in panel (g).}
\end{figure*} 

In the second experiment, we excited spin waves by passing a RF signal ($5\,\mathrm{dBm}$, 7–10$\,$GHz, frequency step $50\,\mathrm{MHz}$, the acquisition time for a single spectrum was $150\,\mathrm{s}$) through the microwave antenna. The spin-wave intensity for each excitation frequency was acquired at two positions: the first position was $x=-0.5\,\mathrm{\upmu m}$ (before the DW) and the second position was $x=0.5\,\mathrm{\upmu m}$ (after the DW). The resulting spectra are shown in Fig. 1(e). Based on these measurements we selected two frequencies ($7.15\,\mathrm{GHz}$ and $9.00\,\mathrm{GHz}$, indicated by red arrows), where we observed an increased spin-wave intensity at both positions, before and after the DW.  

For the selected frequencies we performed intensity- and phase-resolved $\upmu$BLS experiments. The measurements were done for the magnonic  waveguide in a single domain state (SD) and in the DW state (here, DW state means two-domain state separated by the Néel domain wall). For the SD state, we observe only the central waveguide mode [Fig. 2(a)].  The corresponding spin-wave phase ($\Phi$), obtained by the technique described in\cite{PIRRO11, VOGT09},  shows a linear increase in the region -1$\,\mathrm{\upmu m}$ < $x$ < $1\,\mathrm{\upmu m}$ [Fig. 2(d), blue region] as expected for a propagating wave. Outside of this region the phase could not be reconstructed due to either nonlinear spin-wave behavior (closer to the antenna) or insufficient spin-wave signal (farther from the antenna). We determine the wavelength of the excited spin waves from the slope of the fitted line [Fig. 2(d), red line], $\lambda=4.5\pm0.2\,\mathrm{\upmu m}$, which is in agreement with the calculated dispersion relation\cite{KALINIKOS86}. The propagation of the spin waves excited at 7.15$\,$GHz through the DW is shown in Fig. 2(b). The spin-wave propagation pattern before the DW resembles closely the one seen in Fig. 2(a). After passing the DW, the spin waves  are split into two beams. A similar effect of splitting into two beams was observed in a different system by Demidov\cite{DEMIDOV09, DEMIDOV11}. Note, that the spin wave signal is suppressed more strongly at the position of the circular Bloch line, compared to the sides of the domain wall. The phase evolution in the lower spin-wave beam [Fig. 2(e)] exhibits a clear discontinuity; a phase shift of approximately $0.6\uppi$ at the DW position. For the second measured frequency of 9.00$\,$GHz, the 2D spin-wave intensity map looks remarkably different. In this case, the transmitted spin wave is confined to the middle of the waveguide. Also note that the region of the suppressed spin wave signal at the DW position is wider for this frequency than in the case of excitation at $7.15\,\mathrm{GHz}$. The corresponding spin-wave phase, shown in Fig. 2(f), exhibits a phase shift of approximately $0.5\uppi\,$ at the position of the DW ($x=0\,\mathrm{\upmu m}$). 

To further understand this frequency-dependent behavior and phase evolution, we conducted dynamic micromagnetic simulations, using the same material parameters as for the calculation of the static magnetization configuration. Spin waves were excited by a spatially varying field (the field from the rectangular antenna was calculated by FEMM\cite{BALTZIS08}). To prevent reflections at the ends of the simulated waveguide, we implemented regions with increasing damping at both ends of the waveguide. A continuous RF excitation was applied during the initial 20 periods. Then the spatial distribution of the magnetization was sampled for two periods with a time step of $2\,\mathrm{ps}$. The magnetization was transformed to in-plane and out-of-plane components   
%\begin{subequations}
%\begin{align}
%m_\mathrm{oop} &= m_z , \\
%m_\mathrm{ip} &= \left(  \vec{M_0} \left(\vec{r} \right) \times \vec{e_z} \right) \cdot \vec{M} \left( \vec{r}, t \right) ,  
%\end{align}
%\end{subequations}
and the spin-wave phase was calculated using 
%\begin{equation}
%\Phi_\mathrm{SW} = \mathrm{atan}\left( \frac{m_\mathrm{oop}}{m_\mathrm{ip}}\right),
%\end{equation}
%following 
the procedure described in \cite{KOERBER17}. The results of these simulations are shown in Fig. 3(a-c) in the form of the time-averaged squared out-of-plane component of the magnetization. Panel (a) shows the spin-wave propagation in the SD state. Panel (b) shows the spin wave-propagation through the DW at the excitation frequency of $7.15\,$GHz and panel (c) at the frequency of $9.00\,$GHz. We can observe qualitatively the same behavior of the spin wave propagation before and after DW as in the experiment. We do not see the variations of the spin wave suppression within the DW, contrary to what we observed in the experiments. 

Simulated phase maps corresponding to the previously described intensity maps are shown in Fig. 3(d-f). For the waveguide in the SD state we can observe that the phase is almost constant across the $y$ coordinate of the $\upmu$BLS spot [see green dot in Fig. 3(d) and black line in Fig. 3(g)]. On the other hand, the phase after passing the DW is much more distorted across relatively small distance [for $7.15\,\mathrm{GHz}$ see green dot in Fig. 3(e) and red line in Fig. 3(g)] and the same applies to $9.00\,\mathrm{GHz}$ spin wave [see Fig. 3(f), green dot and Fig. 3(g), blue line]. This distortion explains why we were not able to see the linear phase evolution in our experiments. We were probing the spin waves with a $300\,\mathrm{nm}$ $\upmu$BLS spot, and the phase is changing by more than $0.5\,\mathrm{rad}$ within the spot size. 
%For the waveguide in the SD state we can observe spatially coherent phase evolution in the waveguide center; where we were able to measure it with $\mathrm{\upmu}$BLS [see green dot in Fig. 3(d) and black line in Fig. 3(g)]. On the other hand, the phase after passing the DW is strongly spatially incoherent across relatively small distance [for $7.15\,\mathrm{GHz}$ see green dot in Fig. 3(e) and red line in Fig. 3(g)] and the same applies to $9.00\,\mathrm{GHz}$ spin wave [see Fig. 3(f), green dot and Fig. 3(g), blue line]. This spatial phase incoherency explains why we were not able to see the linear phase evolution in our experiments. We were probing the spin waves with a $300\,\mathrm{nm}$ BLS spot, and the phase is changing by more than $0.5\,\mathrm{rad}$ within the spot size. 

\begin{figure}
\includegraphics{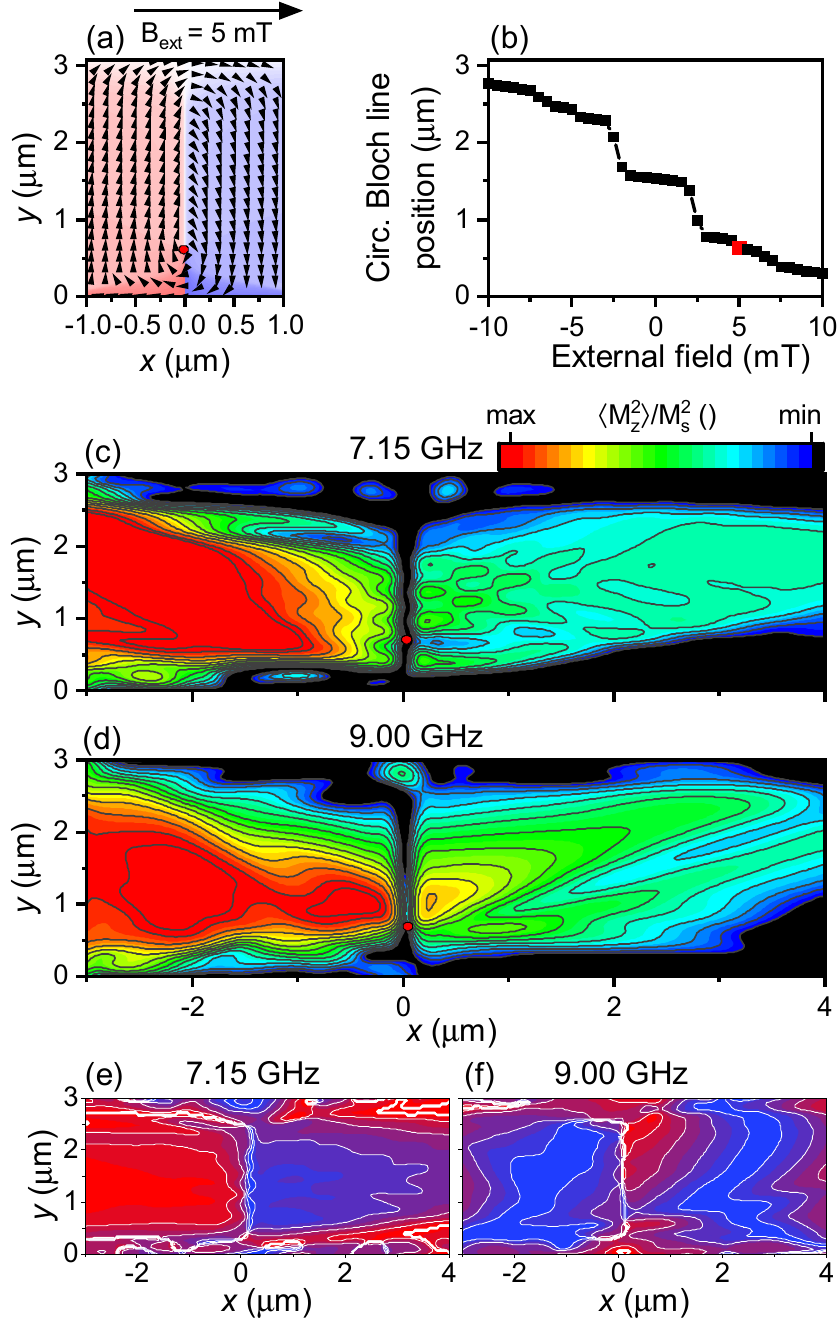}% Here is how to import EPS art
\caption{(a) Magnetization configuration with the external field of $5\,\mathrm{mT}$ applied along the $x$ dimension. The red dots shown in all panels indicate the position of the circular Bloch line at this field. (b) Simulated circular Bloch line displacement as a function of external field in $x$ direction (c-d) Simulated 2D maps of spin wave propagation in an applied field of $5\,\mathrm{mT}$ along the waveguide long axis (c) for frequency 7.15$\,$GHz (d) and for the frequency $9.00\,\mathrm{GHz}$. (e-f) Corresponding 2D maps of the simulated spin wave phase for (e) the frequency $7.15\,\mathrm{GHz}$ and (f) the frequency $9.00\,\mathrm{GHz}$.}
\end{figure}

To check if a change in the internal configuration of the domain wall can be used to modify the spin-wave propagation, we performed a simulation where we applied a small external magnetic field of 5$\,$mT along the $x$ direction. This moves the circular Bloch line in $y$-direction towards the edge of the waveguide [see Fig. 4(a)]. The field dependence of the Bloch line position along the $y$ axis is shown in Fig. 4(b).

This displacement of circular Bloch line changes the transmitted spin-wave profile significantly. In Fig. 4(c) we can observe that for 7.15$\,\mathrm{GHz}$, 
the Bloch line still casts a shadow in the spin-wave intensity; only little space and intensity are left for the lower beam ($y < 0.7\,\upmu\mathrm{m}$). At $f=9\,\mathrm{GHz}$, as in Fig. 3(c), a maximum is seen downstream of the Bloch line, but now displaced sideways, and the wave pattern appears bent. The corresponding phases for the frequencies of 7.15$\,\mathrm{GHz}$ and 9$\,\mathrm{GHz}$ are shown in Fig. 4(e, f).
%the spin waves are not transmitted in the form of two spin-wave beams, but the majority of the intensity passes through the center of the waveguide. For a frequency of 9.00$\,$GHz [Fig. 4(d)], the peak intensity is displaced towards the circular Bloch line position, and the spin wave is “bending” around the Bloch line.

In summary, we stabilized a symmetric Néel DW with a topologically enforced circular Bloch line confined in a magnetic waveguide with imprinted uniaxial anisotropy in the vicinity of a microwave antenna. This configuration allowed us to experimentally observe zero-magnetic-field propagation of DE spin waves through a DW by phase-resolved $\mathrm{\upmu}$BLS. We observed two different regimes of spin-wave propagation appearing at frequencies of 7.15$\,$GHz and 9.00$\,$GHz. In the first regime which was observed at the frequency of 7.15$\,$GHz, spin-wave propagation in the vicinity of the circular Bloch line is suppressed and two spin-wave beams are created. In contrary, at 9.00$\,$GHz spin waves propagate through the circular Bloch line and create a single spin-wave beam. Phase-resolved measurements reveal that spin waves exhibit a phase shift of approximately $0.6\uppi\,$ upon transmission through the DW. 
We observed that the DW spatially distorts the phase of the the spin wave.
%We observed that the DW significantly increases the spatial incoherency of the spin wave phase. 
This effect needs to be taken in account when designing DW-based phase shifters or other devices relying on spin-wave phase manipulation. The other point, which needs to be taken into consideration, is the occurrence of topologically enforced spin structures which are unavoidable in certain geometries. They can limit the performance or functionality of the DW-based device. On the other hand, they can also be an advantage, as in the case where we propose an interesting technique for spin-wave guidance by manipulating the circular Bloch line position in the DW by external magnetic fields. This technique can be used for dynamical turning or blocking spin waves in future magnonic devices.
\vspace{-3pt}
\begin{acknowledgments}
The authors thank to R. Schäfer and O. Fruchart for the discussions on the DW classification. 

This research was supported by CEITEC Nano+ project (ID CZ.02.1.01/0.0/0.0/16013/0001728). 
CzechNanoLab project LM2018110 funded by MEYS CR is gratefully acknowledged for the financial support of the measurement and sample fabrication at CEITEC Nano Research Infrastructure. 
\end{acknowledgments}

The data that support the findings of this study are available from the corresponding author upon reasonable request.

\appendix

%\section{Appendixes}

\nocite{*}
\bibliographystyle{apsrev4-2}
\bibliography{aipsamp}% Produces the bibliography via BibTeX.

\end{document}